\documentclass[aps,prb,twocolumn,showpacs,groupedaddress]{revtex4}

\usepackage{graphicx}

\begin{document}


\title{ Magnetic-field induced superconductor-metal-insulator 
transitions in bismuth metal-graphite }


\author{Masatsugu Suzuki }
\email[]{suzuki@binghamton.edu}
\affiliation{ Department of Physics, State University of New York at 
Binghamton, 
Binghamton, New York 13902-6016}

\author{Itsuko S. Suzuki }
\affiliation{ Department of Physics, State University of New York at 
Binghamton, 
Binghamton, New York 13902-6016}

\author{Robert Lee}
\affiliation{ Department of Physics, State University of New York at 
Binghamton, 
Binghamton, New York 13902-6016}
\author{J\"{u}rgen Walter }
\affiliation{Department of Materials Science and Processing, 
Graduated School of Engineering, Osaka University, 2-1, Yamada-oka, 
Suita, 565-0879, JAPAN}


\date{\today}

\begin{abstract}
Bismuth-metal graphite (MG) has a unique layered 
structure where Bi nanoparticles are encapsulated between adjacent 
sheets of nanographites. The superconductivity below $T_{c}$ 
(= 2.48 K) is due to Bi nanoparticles. The Curie-like susceptibility 
below 30 K is due to conduction electrons localized near zigzag 
edges of nanographites. 
A magnetic-field induced transition from metallic to 
semiconductor-like phase is observed in the in-plane resistivity 
$\rho_{a}$ around $H_{c}$ ($\approx$ 25 kOe) for both $H$$\perp$$c$ 
and $H$$\parallel$$c$ ($c$: $c$ axis). A negative magnetoresistance 
in $\rho_{a}$ for $H$$\perp$$c$ (0$<H\leq$3.5 kOe) and a logarithmic 
divergence in $\rho_{a}$ with decreasing temperature for 
$H$$\parallel$$c$ ($H$ $>$ 40 kOe) suggest the occurrence of 
two-dimensional weak localization effect. 
\end{abstract}


\maketitle



\section{\label{intro}Introduction}
A weak localization theory predicts a logarithmic divergence 
of the resistivity in the two-dimensional (2D) electron systems 
as the temperature ($T$) is lowered.\cite{Abrahams1979} In high-mobility Si 
metal oxide-semiconductor field-effect transistor (MOSFET), 
the in-plane resistivity for a system with an electron density $n$ 
larger than a critical electron density $n_{c}$ decreases with 
decreasing $T$, indicating a metallic behavior.
\cite{Kravchenko1995,Simonian1997,Kravchenko2000} This metallic 
state is completely destroyed by the application of an external 
magnetic field ($H$) applied in the basal plane when $H$ 
is higher than a threshold field $H_{c}$. Such coplanar fields 
only polarize the spins of the electrons, indicating that the 
spin state is significant to the high conductivity of the metallic 
state. The scaling relation of the in-plane resistivity collapses 
into two distinct branches above and below $H_{c}$. Such behaviors 
are very similar to those observed in amorphous ultrathin metal 
films of InO$_{x}$,\cite{Hebard1990} MoGe,\cite{Yazdani1995,Mason1999} 
and Bi,\cite{Markovic1998} which undergo magnetic 
field-induced transitions from superconducting phase to insulating 
phase. 

Bi-metal graphite (MG) constitutes a novel class of materials 
having unique layered structures. This system can be prepared 
from the reduction by Li-diphenylide from an acceptor-type BiCl$_{3}$ 
graphite intercalation compound (GIC) as a precursor material. 
Ideally, the staging structure of Bi-MG would be the same as 
that of BiCl$_{3}$ GIC.\cite{Walter1998,Walter1999a,McRae2001} 
In Bi-MG, Bi atoms would form intercalate 
layers sandwiched between adjacent graphite layers. For the stage-$n$ 
(= 1, 2,$\cdots$) structure, the packages of $n$ graphite layers 
and one Bi layer would be periodically stacked along the $c$ 
axis perpendicular to the graphite basal plane. In BiCl$_{3}$ GIC, 
the ratio of $c$-axis resistivity $\rho_{c}$ to the in-plane 
resistivity $\rho_{a}$ ($\Delta = \rho_{c}/\rho_{a}$) 
at room temperature is (7.4 - 4.9) $\times$ 10$^{3}$,\cite{McRae2001} 
suggesting that 
virtually all $\pi$-electrons are directionally localized; 
i.e., they can move freely along the graphite basal planes, but 
are unable to diffuse across the stack of layers.\cite{Suzuki1996} Such a 
large ratio $\Delta$ is due to the insulating BiCl$_{3}$ layer 
sandwiched between the adjacent graphite layers. There is no 
overlapping of the wave functions over nearest neighbor graphite 
layers. The situation may not drastically change in Bi-MG in 
spite of the fact that BiCl$_{3}$ layer is replaced by metallic 
Bi layer. In fact the $c$-axis resistivity of Bi-MG (= 0.1 $\Omega$cm 
at 298 K) is almost the same as that of BiCl$_{3}$ GIC at the same $T$, 
suggesting that Bi-MG behaves like a quasi 2D conductor.

In this paper we have undertaken an extensive study on the transport 
and magnetic properties of Bi-MG. We show that this compound 
undergoes a superconducting transition at $T_{c}$ = 2.48 K. A 
magnetic-field induced transition from metallic to semiconductor-like 
phase is observed in $\rho_{a}$ 
around $H$ $\approx$ 25 kOe for $H$$\perp$$c$ and $H$$\parallel$$c$ 
($c$: $c$ axis). These results of Bi-MG are compared with 
those of Si MOSFET, and amorphous ultrathin metal films of 
InO$_{x}$,\cite{Hebard1990} 
MoGe,\cite{Yazdani1995,Mason1999} and Bi.\cite{Markovic1998} 

Structural studies of Bi-MG reveal that Bi layers are formed 
of Bi nanoparticles which are encapsulated between adjacent sheets 
of nanographites. The size of nanographites in Bi-MG is much 
smaller than the in-plane coherence size of the graphite layers 
of pristine graphite. The superconductivity is mainly due to 
Bi nanoparticles, while the magnetism is mainly due to nanographites. 
Nanographites are nanometer-sized graphite fragments which represent 
a new class of mesoscopic system intermediate between aromatic 
molecules and extended graphene sheets. Fujita and 
co-workers\cite{Fujita1996,Wakabayashi1998,Wakabayashi1999} 
have theoretically suggested that the electronic structures of 
finite-size graphene sheets depend crucially on the shape of 
their edges. The graphene edge of an arbitrary shape consists 
of two-types of edges, zigzag type and armchair type. The former 
has a trans-polyacetylene type structure and the latter has a 
cis-polyacetylene one. Finite graphite systems having zigzag 
edges exhibit a special edge state. The corresponding energy 
bands are almost flat at the Fermi energy, thereby giving a sharp 
peak in the density of states (DOS) at the Fermi energy. This 
is in contrast to the case of 2D graphene sheet with infinite 
size, where the DOS is zero at the Fermi energy. According to 
Wakabayashi et al.,\cite{Wakabayashi1999} the magnetism of nanographites 
is characterized 
by Pauli paramagnetism and orbital diamagnetism from conduction 
electrons. They have predicted that the Pauli paramagnetic susceptibility 
for the nanographites with zigzag edges shows a Curie-like behavior 
at low temperatures, which is in contrast to a $T$-independent 
Pauli paramagnetism in normal metals. In this paper we show that 
the susceptibility of Bi-MG has a Curie-like behavior at low 
T. This behavior is discussed in the light of the prediction 
by Wakabayashi et al.\cite{Wakabayashi1999}

\section{\label{exp}Experimental procedure}
BiCl$_{3}$ GIC samples as a precursor material, were prepared by 
heating a mixture of highly oriented pyrolytic graphite (HOPG) 
[grade ZYA from Advanced Ceramics, Ohio] and an excess amount 
of BiCl$_{3}$ at 200 $^\circ$C in a ampoule filled with chlorine 
gas at a pressure of 375 Torr.\cite{Walter1998,Walter1999a} 
The reaction was continued 
for three days. It was confirmed from (00$L$) x ray diffraction 
(Rigaku RINT 2000 x-ray powder diffractometer) that the BiCl$_{3}$ 
GIC sample consists of stage-2 as majority phase and stage-3 
and stage-4 as minority phase. The $c$ axis repeat distance is 
13.17 $\pm$ 0.05 $\AA$ for stage-2, 15.85 $\pm$ 0.25 $\AA$ 
for stage-3, and 20.22 $\pm$ 0.25 $\AA$ for stage-4, respectively. 
No Bragg reflection from the pristine graphite is observed.

The synthesis of Bi-MG was made by the reduction by Li-diphenylide 
from BiCl$_{3}$ GIC. BiCl$_{3}$ GIC samples were kept for three 
days in a solution of lithium diphenylide in tetrahydrofuran 
(THF) at room temperature. Then the samples were filtered, rinsed 
by THF, and dried in air. Finally the samples were annealed at 
260 $^\circ$C in a hydrogen gas atmosphere for one day. The 
structure of Bi-MG thus obtained was examined by (00$L$) x-ray 
diffraction, and bright field images and selected-area electron 
diffraction (SAED) (Hitachi H-800 transmission electron microscope) 
operated at 200 kV. The same methods for the structural analysis 
were used for Pd-MG.\cite{Walter1999b,Walter2000a,Walter2000b} 
The (00$L$) x-ray diffraction pattern 
of Bi-MG is much more complicated than that of BiCl$_{3}$ GIC, which 
makes it difficult to calculate the average particle thickness from 
the identity period in Bi-MG. 
Note that graphite reflections appear in Bi-MG, suggesting that 
a part of Bi atoms leaves from the graphite galleries occupied 
by BiCl$_{3}$ intercalate layers in BiCl$_{3}$ GIC during the reduction 
process. Such Bi atoms tend to form multilayered Bi nanoparticles 
in the graphite galleries in Bi-MG. The number of Bi layers in 
possible multilayered structures could not be exactly determined 
at present. 
SAED pattern of Bi-MG consists of polycrystalline diffraction rings, 
suggesting that there are at least four Bi layers in thickness. 
Reflections from Bi and graphite were observed. As listed in Table 
\ref{table:one}, all
the Bi reflections were labeled and attributed to the formation of
rhombohedral Bi, according to the standard ICDD PDF (Card No. 05-0519). 
This result indicates that Bi nanoparticles are crystallized as
rhombohedral Bi phase in Bi-MG. The observed spacings of Bi-MG are 
1-2 \% 
shorter than those of bulk Bi metal. 
The size of Bi 
nanoparticles distribute widely around the average size 110 $\AA$. 
More than 50 \% of Bi nanoparticles has sizes ranging between 
10 and 50 $\AA$. The largest particle size is 750 $\AA$.

\begin{table}
    \caption{\label{table:one} Experimentally observed spacings $d_{exp}$ for Bi-MG
    and $d_{PDF}$ given by powder diffraction file (PDF) 5-0519 for rhombohedral Bi
    [spacegroup: R-3m ($a_{0}$ = 4.546 $\AA$, $c_{0}$ = 11.860 $\AA$ for the notation of the
    hexagonal close-packed structure)].  All reflections can be indexed as
    ($hkl$) reflections of Bi.  Graphite reflections are not included.}
\begin{ruledtabular}
\begin{tabular}{lll}
$d_{exp}$($\AA$) & $d_{PDF}$($\AA$) & ($hkl$)\\
3.24 & 3.28 & 102\\
2.35 & 2.39 & 014\\
1.92 & 1.970 & 113\\
1.71 & 1.868 & 022\\
1.51 & 1.515 & 025\\
1.40 & 1.443 & 212\\
\end{tabular}
\end{ruledtabular}
\end{table}

The measurements of DC magnetization and resistivity were made 
using a SQUID magnetometer (Quantum Design MPMS XL-5) with an 
ultra low field capability and an external device control mode. 
In the present work we used a Bi-MG sample based on HOPG which 
is partially exfoliated. The stoichiometry of C and Bi was not 
determined. The in-plane resistivity $\rho_{a}$ and the $c$ 
axis resistivity $\rho_{c}$ were measured by a conventional 
four-probe method. The sample had a rectangular form with a base 
6.0 $\times$ 1.6 mm$^{2}$ and a height 0.47 mm along the $c$ axis. For the 
measurement of $\rho_{a}$, four thin gold wires (25-$\mu$m 
diameter) that were used as as the current and voltage probes 
were attached to one side of the $c$ surfaces by silver paste 
(4922N, du Pont). For the measurement of $\rho_{c}$, two 
thin gold wires as the current and voltage probes were attached 
to each $c$-surface of the sample. The current ($I$ = 10 
mA for $\rho_{a}$ and 3 mA for $\rho_{c}$) was supplied 
through the current probes by a Keithley type 224, programmable 
DC current source. The voltage $V$ generated across the voltage 
probes was measured by a Keithley 182 nanovoltmeter. The linearity 
of $I$-$V$ characteristics was confirmed for the measurements 
of $\rho_{a}$ and $\rho_{c}$.

\section{\label{result}Result and discussion}
\subsection{\label{resultA}Meissner effect due to Bi nanoparticles}

\begin{figure}
\includegraphics[width=8.0cm]{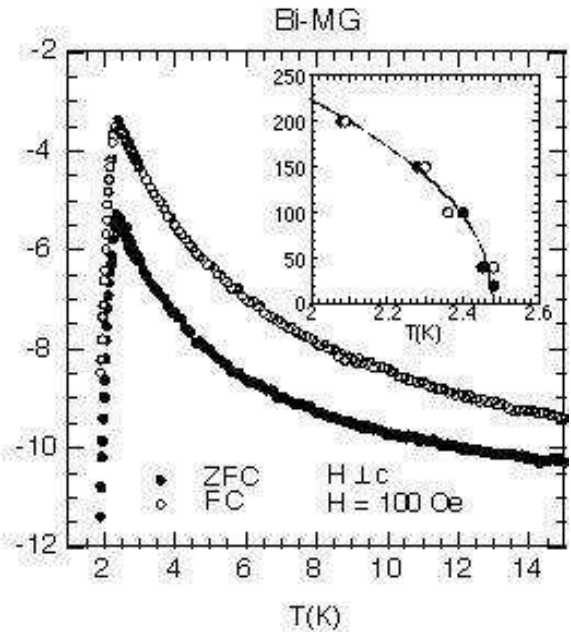}%
\caption{\label{fig:one} $T$ dependence of ZFC and FC 
susceptibilities of Bi-MG based on HOPG. $H$ = 100 Oe. 
$H$$\perp$$c$ ($c$: $c$ axis). The inset shows the magnetic 
phase diagram of $H$ vs $T$ for 
$H$$\perp$$c$, where the peak temperatures 
of ZFC and FC susceptibilities are plotted as a function of 
$H$. The solid line denotes the least-squares fit of the ZFC 
data ({\Large $\bullet$}) to Eq.(\ref{eq:one}).}
\end{figure}

\begin{figure}
\includegraphics[width=8.0cm]{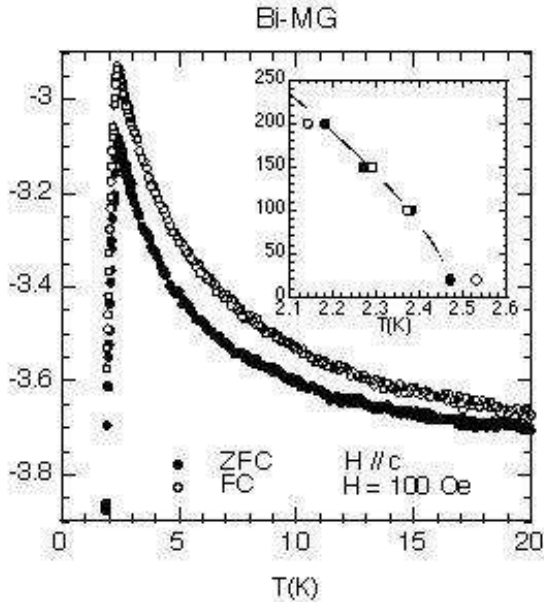}%
\caption{\label{fig:two} $T$ dependence of ZFC and FC 
susceptibilities of Bi-MG. $H$ = 100 Oe. 
$H$$\parallel$$c$ ($c$: $c$ axis). The inset shows the magnetic phase diagram of 
$H$ vs $T$ for $H$$\parallel$$c$. The solid line 
denotes the least-squares fitting curve for the ZFC data ({\Large $\bullet$}) 
to Eq.(\ref{eq:one}).}
\end{figure}

We measured the DC magnetization of Bi-MG. After the sample was 
cooled from 298 to 1.9 K at $H$ = 0, the measurement of zero 
field cooled (ZFC) magnetization $M_{ZFC}$ was made with increasing $T$ 
from 1.9 to 15 K in the presence of $H$. The sample was kept 
for 20 minutes at 50 K. Then the sample was cooled from 50 
to 15 K. The measurement of field-cooled (FC) magnetization $M_{FC}$ 
was made with decreasing $T$ from 15 to 1.9 K in the presence 
of the same $H$. For convenience, hereafter, the direction 
of $H$ in the $c$ plane is denoted as $H$$\perp$$c$ ($c$: $c$ axis). 
Figures \ref{fig:one} and \ref{fig:two} show typical examples of the $T$ dependence 
of the susceptibility $\chi_{ZFC}$ (=$M_{ZFC}/H$) and $\chi_{FC}$ 
(=$M_{FC}/H$) at $H$ = 100 Oe for $H$$\perp$$c$ and $H$$\parallel$$c$, 
respectively. Both $\chi_{FC}$ and $\chi_{ZFC}$ show 
a sharp peak around 2.5 K at $H$ = 20 Oe, which results from 
the competition between the diamagnetic susceptibility due to 
the Meissner effect and the Curie-like susceptibility (which 
will be described later). The peak shifts to the low $T$ side 
with increasing $H$. No peak is observed above 1.9 K for $H\geq$ 250 
Oe. For convenience, the peak temperature is defined as a critical 
temperature $T_{c}(H)$. The peak temperature $T_{c}(H)$ 
for $\chi_{FC}$ is almost the same as that for $\chi_{ZFC}$. 
In the insets of Figs. \ref{fig:one} and \ref{fig:two}, we show the $H$-$T$ magnetic 
phase diagram for Bi-MG for $H$$\perp$$c$ and $H$$\parallel$$c$, 
respectively. We find that the data of $H$ vs $T$ (ZFC susceptibility) 
for $H$$\perp$$c$ and $H$$\parallel$$c$ are well described by
\begin{equation}
H=H_{0}(\frac{T_{c} -T}{T_{c} } )^{\beta},
\label{eq:one}
\end{equation}
where $H_{0}$ is the critical field at $T$ = 0 K ($H_{0}$ = $H_{c2}$ for
$H$$\perp$$c$ and $H_{0}$ = $H_{c3}$ for $H$$\parallel$$c$), $\beta$ is an
exponent, and $T_{c}$ is a critical temperature at $H$ = 0.  The field
$H_{c3}$ is a field to nucleate a small superconducting region near the
sample surface.  The least squares fits of the data of $H$ vs $T$ (ZFC
susceptibility) yield the values of $H_{c2}$ = 489.4 $\pm$ 0.5 Oe, $\beta$
= 0.48 $\pm$ 0.02, $T_{c}$ = 2.48 $\pm$ 0.06 K for $H$$\perp$$c$, and
$H_{c3}$ = 838 $\pm$ 1 Oe, $\beta$ = 0.69 $\pm$ 0.02, $T_{c}$ = 2.48 $\pm$
0.02 K for $H$$\parallel$$c$.
Here we note that the superconductivity is observed in ultrathin films of
amorphous Bi grown in top of layer of amorphous Ge (thickness, 6 - 10
$\AA$).  Haviland et al.\cite{Haviland1989} have reported that the
superconductivity occurs for the thickness of Bi films, $d$ = 6.73 - 74.27
$\AA$.  The critical temperature $T_{c}$ decreases with decreasing $d$:
$T_{c}$ = 5.6 K for $d$ = 74.27 $\AA$ and $T_{c}$ $\approx$ 0.8 K for $d$
$\approx$ 6.73 $\AA$.  Markovic et al.\cite{Markovic1998} have shown that
$T_{c}$ decreases with decreasing $d$ for $d>d_{c}$ ($d_{c}$ = 12.2 $\AA$):
$T_{c}$ = 0.5 K for $d$ = 15 $\AA$.  Weitzel and Micklitz\cite{Weitzel}
have reported surface superconductivity in granular films built from
well-defined rhombohedral Bi clusters (mean size = 38 $\AA$) embedded in
difference matrices (Kr, Xe, Ge) or with H$_{2}$ or O$_{2}$ gas adsorbed on
the cluster surface.  The critical temperature $T_{c}$, which is dependent
on matrices and gases used, is between 2 and 6 K. This value of $T_{c}$ is
on the same order as that of Bi-MG.
The exponent $\beta$ for $H$$\perp$$c$ 
is very close to that ($\beta$ = 0.5) predicted for homogeneous 
system of isolated superconducting grains.\cite{Deutscher1980} 
The ratio $H_{c3}$/$H_{c2}$ 
is calculated as 1.71, which is very close to the predicted value 
1.695.\cite{SaintJames1963} Similar behavior is observed in the critical fields 
for $H$$\perp$$c$ and $H$$\parallel$$c$ in stage-1 K GIC 
(KC$_{8}$).\cite{Tanuma1992} 
The coherence length $\xi$ is estimated as $\xi$ 
= 820 $\AA$ from the value of $H_{c2}$ using the relation $H_{c2}$ 
= $\Phi_{0}$/(2$\pi \xi ^{2}$), where $\Phi_{0}$ 
(= 2.0678 $\times$ 10$^{-7}$ Gauss cm$^{2}$) is a fluxoid. 
The coherence length $\xi$ 
is much larger than the size of islands (110 $\AA$ in average).

\begin{figure}
\includegraphics[width=8.0cm]{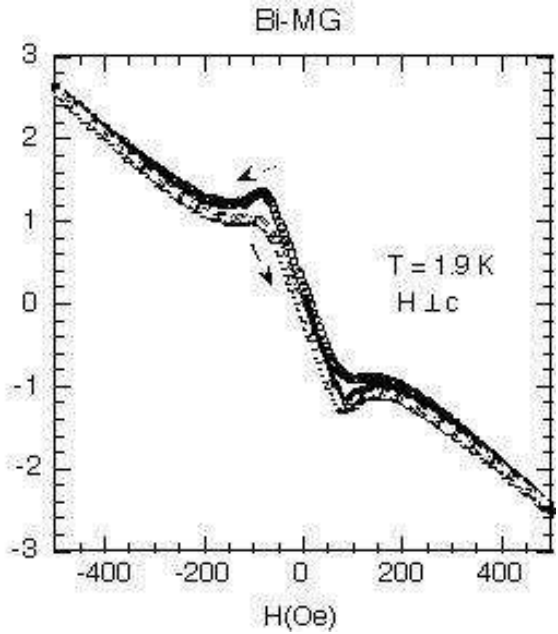}%
\caption{\label{fig:three} Hysteresis loop of DC magnetization for 
Bi-MG. $H$$\perp$$c$. $T$ = 1.9 K. 
The measurement was made with increasing $H$ from 0 to 500 Oe 
(denoted by {\Large $\bullet$}), with decreasing $H$ from 
500 Oe to -500 Oe ({\Large $\circ$}), and with increasing $H$ 
from -500 Oe to 500 Oe ($\triangle$).}
\end{figure}

\begin{figure}
\includegraphics[width=8.0cm]{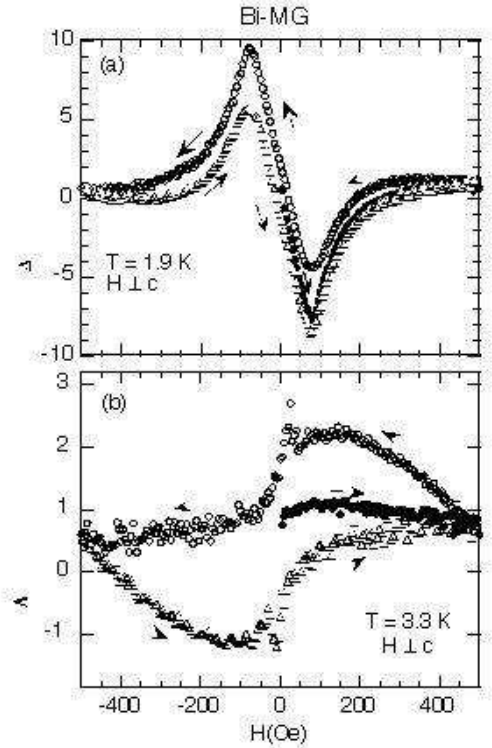}%
\caption{\label{fig:four} Hysteresis loop of DC magnetization minus a 
diamagnetic background (= $\chi$$_{d}$$H$) 
for Bi-MG. $H$$\perp$$c$. (a) $T$ = 
1.9 K. $\chi$$_{d}$ = -5.143 $\times$ 10$^{-7}$ emu/gOe. 
(b) $T$ = 3.3 K. $\chi$$_{d}$ = -7.317 $\times$ 
10$^{-7}$ emu/gOe.}
\end{figure}

Figure \ref{fig:three} shows the hysteresis loop of the 
magnetization $M_{a}(H)$ 
for $H$$\perp$$c$ at $T$ = 1.9 K. The sample was cooled 
from 298 K to 1.9 K at $H$ = 0. The magnetization $M_{a}(H)$ 
was measured with increasing $H$ from 0 to 500 Oe, with decreasing $H$ 
from 500 Oe to -500 Oe, and with increasing $H$ from -500 Oe 
to 500 Oe. The magnetization consists of superconductivity contribution 
and diamagnetic background. In Fig. \ref{fig:four}(a) we show the hysteresis 
loop of $\Delta M_{a}(H)$ that is defined by $M_{a}(H)$ 
at 1.9 K minus a diamagnetic background given by $\chi_{d}H$ 
with $\chi_{d}$ = -5.143 $\times$ 10$^{-7}$ emu/gOe. A hysteresis 
loop characteristic to a type-II superconductor is observed with 
a lower critical field $H_{c1}$ (= 80 Oe). In contrast, the magnetization 
hysteresis loop$M_{a}(H)$ at $T$ = 3.3 K is very different 
from that at 1.9 K. It seems that there is neither local minimum 
nor local maximum. In Fig. \ref{fig:four}(b) we show the hysteresis 
loop of $\Delta M_{a}(H)$ 
which is defined by at $M_{a}(H)$ at 3.3 K minus a diamagnetic 
background given by $\chi_{d}H$ with $\chi_{d}$ 
= -7.317 $\times$ 10$^{-7}$ emu/gOe. As $H$ decreases, a trapped magnetic 
flux, corresponding to a paramagnetic moment $M_{r}$ ($\approx$ 
2.7 $\times$ 10$^{-5}$ emu/g) remains in the sample. With further cycling 
of $H$ from -500 to 500 Oe a characteristic hysteresis loop 
is observed. 

We also measured the hysteresis loop of the magnetization $M_{c}(H)$ 
for $H$$\parallel$$c$ at $T$ = 1.9 K. The difference $\Delta M_{c}$ 
is defined by $M_{c}(H)$ at 1.9 K minus a diamagnetic background 
given by $\chi_{d}H$ with $\chi_{d}$ = -2.980 $\times$ 
10$^{-6}$ emu/gOe. The hysteresis loop of $\Delta M_{c}$ 
for $H$$\parallel$$c$ is very similar to the corresponding data $\Delta M_{a}$ 
for $H$$\perp$$c$. The value of $H_{c1}$ (= 75 Oe) for $H$$\parallel$$c$ 
is a little lower than that for $H$$\perp$$c$. The peak 
value of $\Delta M_{c}$ is almost the same as that of $\Delta M_{a}$. 
The hysteresis loop of $\Delta M_{c}$ at 1.9 K exhibits 
a nearly reversible behavior. In bulk samples this could be explained 
in terms of a lack of structural defects to provide pinning sites 
in the vortex state ($H_{c1} < H < H_{c2}$). In Bi-MG, 
however, the size of islands is much shorter than the coherence 
length $\xi$. Thus the pinning effect is not relevant 
for these islands. In contrast, the hysteresis loop of $\Delta M_{a}$ 
at 1.9 K exhibits some hysteretic behavior. Note that similar 
magnetization curves are observed in a superconductor TaC nanoparticles 
which are synthesized using a vapor-solid reaction path starting 
with carbon nanotube precursor.\cite{Fukunaga1998}

\subsection{\label{resultB}Curie-like susceptibility due to nanographites}

\begin{figure}
\includegraphics[width=8.0cm]{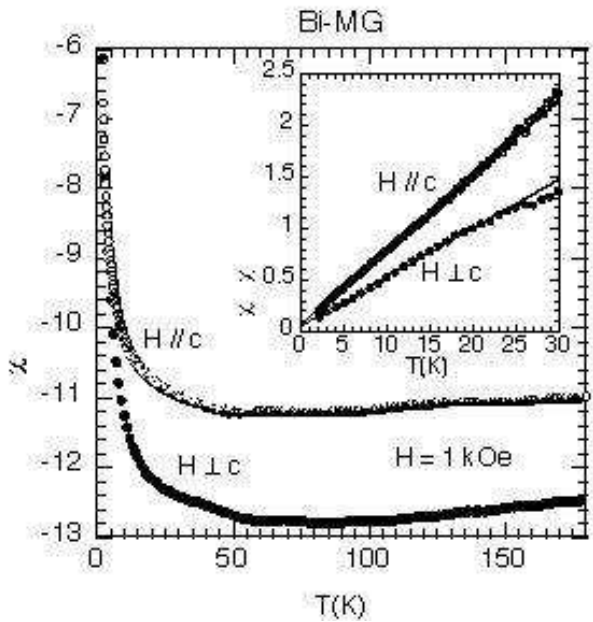}
\caption{\label{fig:five} $T$ dependence of FC 
susceptibilities for Bi-MG. $H$ = 1 kOe. 
$H$$\perp$$c$ and $H$$\parallel$$c$. The inset 
shows the reciprocal susceptibility ($\chi-\chi_{0}$)$^{-1}$ as a function of $T$. 
The solid lines are described by Eq.(\ref{eq:two}) with parameters 
given in the text. }
\end{figure}

\begin{figure}
\includegraphics[width=8.0cm]{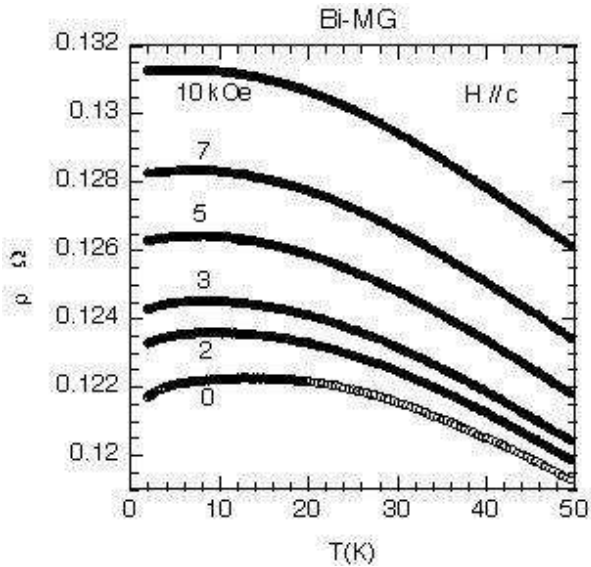}
\caption{\label{fig:six}$T$ depedence of $c$-axis resistivity $\rho_{c}$($T$,$H$) 
for Bi-MG at various $H$. $H$$\parallel$$c$. $I$$\parallel$$H$.}
\end{figure}

Figure \ref{fig:five} shows typical data of $\chi_{FC}$ for $H$ = 1 
kOe for $H$$\perp$$c$ and $H$$\parallel$$c$. The susceptibility 
is negative at high $T$ and drastically increases with decreasing $T$. 
The least squares fit of the data ($\chi_{FC}$ vs $T$) 
for 1.9 $\leq T \leq$ 30 K to the Curie-Weiss law
\begin{equation}
\chi =\chi _{0} +\frac{C}{T-\Theta },
\label{eq:two}
\end{equation}
yields $\Theta$ = -0.86 $\pm$ 0.05 K, $C$ = (1.999 
$\pm$ 0.029) $\times$ 10$^{-6}$ emu K/g, $\chi_{0}$ = (-1.307 $\pm$ 
0.001) $\times$ 10$^{-6}$ emu/g for $H$$\perp$$c$, and $\Theta$ 
= -1.13 $\pm$ 0.02 K, $C$ = (1.460 $\pm$ 0.008) $\times$ 10$^{-6}$ 
emu K/g, $\chi_{0}$ = (-1.146 $\pm$ 0.001) $\times$ 10$^{-6}$ emu/g 
for $H$$\parallel$$c$. The value of $\Theta$ is very close to 
zero, showing a Curie-like law. In the inset of Fig. \ref{fig:five} we show 
the reciprocal susceptibility ($\chi -\chi_{0})^{-1}$ 
as a function of $T$. The negative value $\chi_{0}$ is 
from the orbital diamagnetic susceptibility. There is a crossover 
from the high-temperature diamagnetic susceptibility to the low-temperature 
Curie-like susceptibility around 50 K. We assume that the susceptibility 
of Bi-MG at 100 K corresponds to the diamagnetic susceptibility 
since the Pauli susceptibility is positive and is nearly equal 
to zero at 100 K. From Fig. \ref{fig:five} the diamagnetic susceptibility for 
$H$$\parallel$$c$ 
and $H$$\perp$$c$ is estimated as $\chi_{c} \approx$ 
-1.1 $\times$ 10$^{-6}$ emu/g and $\chi_{a} \approx$ - 1.3 $\times$ 10$^{-6}$ 
emu/g, which are almost isotropic. These values are in contrast 
with the susceptibility of HOPG, which is very anisotropic: $\chi_{c}$ 
= -25.86 $\times$ 10$^{-6}$ emu/g and $\chi_{a}$ = -1.06 $\times$ 10$^{-7}$ emu/g 
at $H$ = 1 kOe. The absolute value of $\chi_{c}$ in Bi-MG 
is much smaller than that in HOPG, while the values of $\chi_{a}$ 
are on the same order for both systems.

Here it is interesting to compare our data of susceptibility 
with that of nanographites prepared by heat treating diamond 
nanoparticles\cite{Anderssen1998} and a disorder network of nanographites in 
activated carbon fiber (ACF).\cite{Shibayama2000} The $T$ dependence of the 
susceptibility for these compounds is similar to that of Bi-MG. 
In particular, the values of $\Theta$, $C$, and $\chi_{0}$ 
for the ACF are on the same order as those for Bi-MG, where $\Theta$ 
$\approx$ -2 K, $C$ = 1.21 $\times$ 10$^{-6}$ emu K/g, and $\chi_{0}$ 
= -1.36 $\times$ 10$^{-6}$ emu/g for ACF prepared at the heat treatment temperature 
HTT = 1500 $^\circ$C.\cite{Shibayama2000} Here we assume that the 
spin $S$ 
is 1/2 and the Land\'{e} $g$-factor is 2 since the spins are 
associated with carbon materials. The number of spin density 
per 1 g of Bi-MG is estimated as (4.7 - 6.4) $\times$ 10$^{18}$/g, which 
is comparable with 3.9 $\times$ 10$^{18}$/g for the ACF with HTT = 1500 
$^\circ$C.\cite{Shibayama2000}

It has been theoretically predicted by 
Wakabayashi et al.\cite{Wakabayashi1999} 
that the Pauli susceptibility exhibits a Curie-like behavior 
in the nanographites with zigzag edges, because of the DOS which 
is sharply peaked at the Fermi energy. However, qualitatively 
we think that the Curie-like behavior at low T in Bi-MG is due 
to the conduction electrons localized around the zigzag edges 
of nanographites, which have local magnetic moments (spin $S$ 
= 1/2 and a Land\'{e} $g$-factor $g$ = 2). The origin of spin 
polarization in nanographites with zigzag edges has been discussed 
by Fujita et al.\cite{Fujita1996} using the Hamiltonian that consists of the 
on-site Coulomb repulsive interaction ($U$) when the site is 
occupied by two electrons, and the electron transfer integral 
between the nearest sites ($t$). They have shown that a ferrimagnetic 
spin polarization emerges on the edge carbons even for weak $U/t$ 
$\approx$ 0.1.

The orbital diamagnetic susceptibility is very sensitive to the 
size and edge shapes of nanographites. The orbital diamagnetic 
susceptibility is almost isotropic in Bi-MG, while it is very 
anisotropic in pristine graphite with infinite size. In Bi-MG 
the orbital cyclotron motion of electrons in the presence of $H$ 
($H$$\parallel$$c$) is greatly suppressed. This result is in good agreement 
with the prediction by Wakabayashi et al.\cite{Wakabayashi1999} In ribbon-shaped 
nanographites with zigzag edges, the magnitude of the diamagnetic 
susceptibility decreases as the ribbon width decreases. The flow 
of the orbital diamagnetic ring current significantly depends 
on the lattice topology near the graphite edge.

\subsection{\label{resultC}$c$-axis and in-plane resistivities}

\begin{figure}
\includegraphics[width=8.0cm]{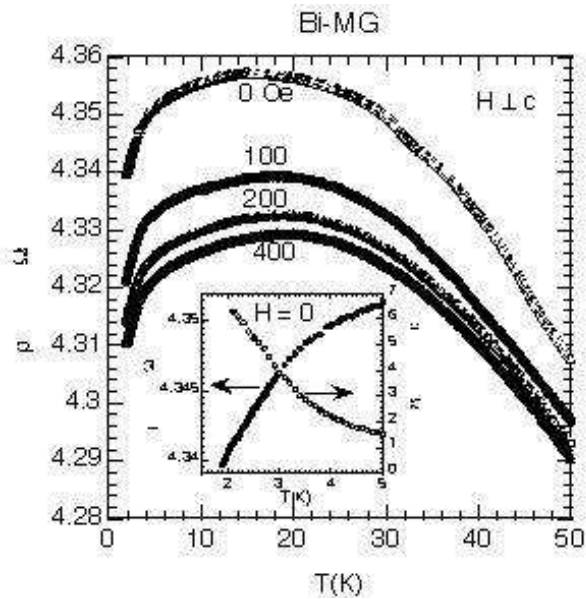}%
\caption{\label{fig:seven} $T$ dependence of in-plane 
resistivity $\rho$$_{a}$($T$,$H$) 
for Bi-MG at $H$ = 0, 100, 200, and 400 Oe. 
$H$$\perp$$c$. $I$$\perp$$H$. The inset shows the $T$ dependence of 
$\rho$$_{a}$($T$,$H$ = 0) ({\Large $\bullet$}) and d$\rho$$_{a}$($T$, $H$ = 
0)/d$T$ ({\Large $\circ$}) at low 
$T$.}
\end{figure}

The $c$-axis resistivity $\rho_{c}$ of Bi-MG was measured 
using the four-probe method. After the sample was cooled from 
298 K to 1.9 K at $H$ = 0, the $c$-axis resistivity $\rho_{c}$ 
was measured with increasing $T$ from 1.9 K to 50 K without 
and with $H$ [$H$$\parallel$$c$ ($c$: $c$ axis)], where the current direction is 
parallel to the field direction. This resistivity is denoted 
as the longitudinal magnetoresistance. Figure \ref{fig:six} shows 
the $T$ 
dependence of $\rho_{c}$ for Bi-MG at $H$ = 0 - 10 kOe. 
The resistivity $\rho_{c}$ for Bi-MG shows a very broad 
peak around 14 K at $H$ = 0, in contrast to $\rho_{c}$ 
for HOPG exhibiting a peak around 40 K.\cite{Suzuki1996} This peak shifts 
to the low $T$ side with increasing $H$. For $H \geq$ 20 
kOe, $\rho_{c}$ shows a semiconductor-like behavior: it 
decreases with increasing $T$. Note that $\rho_{c}$ shows 
a positive magnetoresistance: it increases with increasing $H$ 
at any fixed $T$ between 1.9 and 50 K. The value of $\rho_{c}$ 
at $T$ = 298 K and $H$ = 0 is 0.11 $\Omega$cm, which is 
on the same order as that of HOPG 
($\rho_{c}$ = 0.096 $\Omega$cm)\cite{Suzuki1996} 
and BiCl$_{3}$ GIC ($\rho_{c}$ = 0.2 $\Omega$cm),\cite{McRae2001} leading 
to a mean free path less than 1 $\AA$ according to the Drude formula. 
This result suggests that there is no overlapping over nearest-neighbor 
layers along the $c$ axis. The \textit{c}-axis conduction can occur 
via the hopping of carriers between layers.

The in-plane resistivity $\rho_{a}$ of Bi-MG was also measured 
using the four-probe method. After the sample was cooled from 
298 K to 1.9 K at $H$ = 0, the in-plane resistivity $\rho_{a}$ 
was measured with increasing $T$ from 1.9 K to 50 K without 
and with $H$ ($H$$\perp$$c$), where the field direction 
is perpendicular to the current direction in the $c$ plane. 
This resistivity is denoted as the transverse magnetoresistance. 
The resistivity ratio $\Delta$ (= $\rho_{c}$/$\rho_{a}$) 
is estimated as 30 at 298 K using the measured $\rho_{a}$. 
However, the actual value of $\Delta$ is considered to 
be much larger than 30 because of possible contribution of $\rho_{c}$ 
to the measured $\rho_{a}$. Figure \ref{fig:seven} shows the $T$ dependence 
of $\rho_{a}$ of Bi-MG at $H$ = 0 - 400 Oe. The zero-field 
($H$ = 0) in-plane resistivity increases with increasing $T$ 
at low $T$, showing a metallic behavior. It has a maximum around 
15 K, and it decreases with further increasing $T$, showing 
a semiconductor-like behavior. Note that the value of $\rho_{a}$ 
for Bi-MG (= 4.2 m$\Omega$cm) is much larger than that of BiCl$_{3}$ 
GIC (= 27 $\mu \Omega$cm at 298 K). We find that the $T$ 
dependence of $\rho_{a}$ for Bi-MG is very similar to that 
of grafoil: they even have similar magnitudes. The grafoil is 
a pyrolytic graphite with a polycrystalline structures formed 
of many domains. According to Koike et al.,\cite{Koike1985} $\rho_{a}$ of 
grafoil (grade GTA, Union carbide) shows a semiconductor-like 
behavior, while $\rho_{a}$ of HOPG (Union Carbide) shows 
a metallic behavior. This result suggests that the semiconductor-like 
behavior in Bi-MG may arise from nanographites where the degree 
of disorder is greatly enhanced. Such a semiconductor-like behavior 
is observed at least below 25 kOe.

In the inset of Fig. \ref{fig:seven} we show the detail of $\rho_{a}$ 
at $H$ 
= 0 near $T_{c}$ = 2.48 K. No drastic decrease in $\rho_{a}$ 
below $T_{c}$ is observed with decreasing $T$, while the $T$-derivative 
$d \rho_{a}/dT$ gradually decreases with increasing $T$ 
around $T_{c}$. The causes for the finite value of $\rho_{a}$ 
below $T_{c}$ in spite of the superconducting phase are as follows. 
(i) The sample used in the present work is an exfoliated Bi-MG 
based on HOPG. The sample surface is not flat partly because 
of cracks generated in the basal plane. Since the current path 
is not always located on the same layer, the contribution of 
large $\rho_{c}$ to the observed $\rho_{a}$ is not 
negligibly small. Another possibility is the local superconductivity 
in isolated islands. In such a system there is competition between 
the Josephson coupling ($E_{J}$) and the charging energy ($E_{U}$) 
between superconducting islands.\cite{Zwerger1991} For $E_{J} \gg E_{U}$, 
the Cooper pairs are delocalized leading to a superconducting 
state with vanishing resistivity. For $E_{J} \ll E_{U}$, 
the pairs will be localized and the transport is possible only 
by thermal activation leading to insulating behavior at $T$ 
= 0 K. The resistivity varies with $T$ as $\rho \approx$ 
exp($T_{a}/T$), where $T_{a}$ is related to the activation 
energy. In this model, the resistivity should increase with decreasing $T$, 
which contradicts with our result. Liu et al.\cite{Liu1992} have reported 
an unusual $T$ dependence of resistivity at low $H$ in ultrathin 
superconducting films of Pb, Al, and Bi. The resistivity varies 
with $T$ as $\rho \approx$ exp($T/T_{0}$) at 
low $T$, where $T_{0}$ is a characteristic temperature. In 
Bi-MG, $d\rho/dT$ decreases with increasing $T$ 
around $T_{c}$. In superconducting thin films, $d\rho/dT$ 
[$\approx$ exp($T/T_{0}$)/$T_{0}$] increases with increasing $T$. 
Therefore, the $T$ dependence of $\rho_{a}$ in Bi-MG is 
not the case of $\rho \approx$ exp($T/T_{0}$).

\subsection{\label{resultD}2D Weak localization effect}

\begin{figure}
\includegraphics[width=8.0cm]{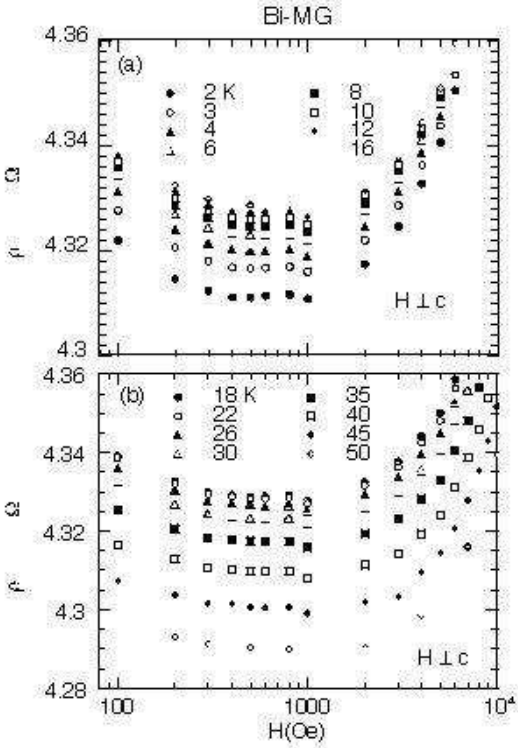}%
\caption{\label{fig:eight} $H$ dependence of 
$\rho$$_{a}$($T$,$H$) for Bi-MG at 
fixed $T$ (2 $\leq$ $T$ $\leq$ 50 K). 
$H$$\perp$$c$.}
\end{figure}

As shown in Fig. \ref{fig:seven}, the in-plane resistivity $\rho_{a}$ slightly 
decreases with increasing $H$ at the same $T$ (at least for 
1.9 $\leq T \leq$ 50 K), indicating a negative magentoresistance 
(NMR). Figures \ref{fig:eight}(a) and (b) show the $H$ dependence of $\rho_{a}$ 
for Bi-MG with 2 $\leq T \leq$ 50 K, where $H$$\perp$$c$ 
and the field direction is perpendicular to the current direction. 
For each $T$, $\rho_{a}$ decreases with increasing $H$ 
at low $H$ ($H \geq$ 100 Oe), exhibits a local minimum 
around $H$ = 2.5 kOe, and increases with further increasing $H$. 
The sign of the difference $\Delta \rho_{a}$ 
[= $\rho_{a}(T,H)-\rho_{a}(T,H=0)$] 
is negative for 0 $\leq H \leq$ 3.5 kOe due to the possible 
2D weak localization effect (WLE).\cite{Hikami1980} It changes from negative 
to positive at $H$ = 3.5 kOe. The resistivity $\rho_{a}$ 
drastically increases with further increasing $H$, as is observed 
in compensated metals such as bulk Bi.\cite{Suzuki1978} The $H$ dependence 
of $\rho_{a}$ for 3 kOe $\leq H \leq$ 45 kOe can 
be well described by $\rho_{a}$ = $\rho_{0}$ + $\rho_{1}H$ 
+ $\rho_{2}H^{2}$, where $\rho_{0}$, $\rho_{1}$, 
and $\rho_{2}$ are constants: $\rho_{0}$ = 4.290 m$\Omega$cm, $\rho_{1}$ 
= (9.85 $\pm$ 0.11) $\times$ 10$^{-3}$ $\mu \Omega$cm/Oe, and $\rho_{2}$ 
= (4.56 $\pm$ 0.23) $\times$ 10$^{-8}$ $\mu \Omega$cm/Oe$^{2}$ at 1.9 
K. The linear term ($\rho_{1}H$) is dominant compared 
to the squared-power term ($\rho_{1}H^{2}$) for $H \ll H_{1}$, 
where $H_{1}$ = $\rho_{1}$/$\rho_{2}$ = 216 kOe at 
1.9 K. For comparison, we also measured the $H$ dependence of $\rho_{a}$ 
in Bi-MG for 1.9 $\leq T \leq$ 30 K, where $H$$\parallel$$c$. 
At each $T$, $\rho_{a}$ increases with increasing $H$. 
Thus the sign of $\Delta \rho_{a}$ is always positive 
for 0 $<H\leq$ 47.5 kOe and 1.9 $\leq T \leq$ 50 
K, indicating an anti-localization effect which is similar to 
that observed in Bi thin films\cite{Komori1983} with strong spin-orbit interactions.

The $T$ dependence of $\rho_{a}$ at $H$ = 45 kOe for $H$$\perp$$c$ 
is described by 
\begin{equation}
\rho _{a} =a_{0} -a_{1} \ln (T),
\label{eq:three}
\end{equation}
for 1.9 $\leq T\leq$ 4.3 K, where $a_{0}$ = (4.8261 $\pm$ 
0.0002) m$\Omega$cm and $a_{1}$ = (3.79 $\pm$ 0.19) $\times$ 10$^{-3}$ 
m$\Omega$cm. As shown in previous papers,\cite{Suzuki1996,Suzuki2000} the theory 
of the 2D WLE predicts that the following relation is valid for 
the ratio $a_{2}$/$a_{1}$,
\begin{equation}
\frac{a_{1} }{a_{0} } =\frac{e^{2} }{2\pi ^{2} \hbar } 
\frac{A}{\sigma_{2D}^{0} },
\label{eq:four}
\end{equation}
with $A=\alpha p+\gamma$, where $e^{2}/(2\pi^{2}\hbar$) 
= 1.23314 $\times$ 10$^{-5}$ $\Omega^{-1}$ and $\sigma_{2D}^{0}$ is the 
in-plane conductivity defined by $I_{c}$/$\rho_{a}^{0}$ ($\rho_{a}^{0}$ 
is the in-plane resistivity and $I_{c}$ is the $c$ axis repeat 
distance). In the parameter $A$, $\alpha$ is nearly equal 
to unity, $p$ is the exponent of the inelastic life time $\tau_{\epsilon}$ 
of the conduction electrons ($\tau_{\epsilon} \approx T^{-p}$), 
and $\gamma$ is a numerical factor giving a measure of the 
screening by other carriers. For convenience, here we use the 
value of $\sigma_{2D}^{0}$ (= 1.89 $\times$ 10$^{-2}$ $\Omega^{-1}$) for 
kish graphite ($I_{c}$ = 3.35 $\AA$ and $\rho_{a}^{0}$ = 1.77 
$\mu \Omega$cm at $T$ = 4.2 K) obtained by Koike et al.,\cite{Koike1985} 
instead of the corresponding data for Bi-MG. 
A kish graphite is a high-quality single crystal of graphite which is
deposited on walls of blast furnaces for steel production.
Using $a_{1}/a_{0}$ 
= 7.853 $\times$ 10$^{-4}$ the parameter $A$ is estimated as $A$ = 1.20, 
which is comparable with 1.14 for stage-4 MoCl$_{5}$ GIC.\cite{Suzuki2000} These 
results suggest that the logarithmic behavior of $\rho_{a}$ 
can be explained in terms of the 2D WLE.

\subsection{\label{resultE}Field induced metal-semiconductor transition}

\begin{figure}
\includegraphics*[bb=50 50 425 750,clip,width=8.0cm]{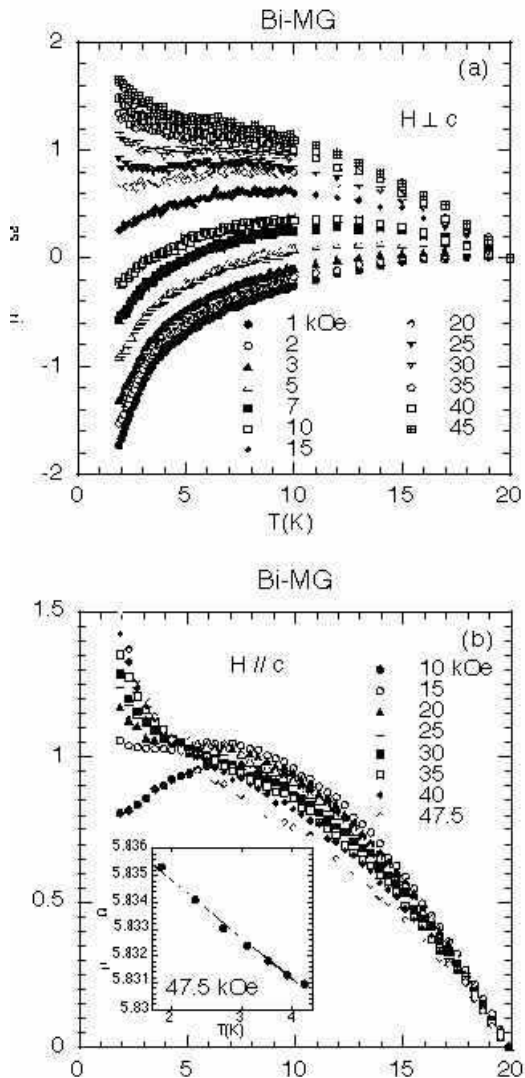}
\caption{\label{fig:nine} $T$ dependence of the difference $\mu$$_{a}$ 
between $\rho _{a} (T,H)$ and $\rho _{a} (T=20K,H)$ [$\mu _{a} =\rho 
_{a} (T,H)-\rho _{a} (T=20K,H)$] for Bi-MG with various $H$ 
(1 $\leq$ $H$ $\leq$ 45 kOe). 
$I$$\perp$$H$. (a) 
$H$$\perp$$c$. (b) $H$$\parallel$$c$. 
The inset of (b) shows the $T$ dependence of $\rho _{a} (T,H)$ 
at $H$ = 47.5 kOe. The solid line denotes the fitting curve of 
the data to Eq.(\ref{eq:three}).}
\end{figure}

In Fig. \ref{fig:nine}(a) we show the $T$ dependence of the 
difference ($\mu_{a}$) 
between $\rho_{a}$ at $T$ and that at 20 K for various $H$ 
($H$$\perp$$c$): $\mu_{a} = \rho_{a}(T,H) 
- \rho_{a}(T=20 K,H)$. The $T$ dependence of $\mu_{a}$ 
below 10 K is dependent on $H$. The difference $\mu_{a}$ 
shows a metallic behavior ($d\mu_{a}/dT>$ 0) below 
15 - 20 kOe, while it shows a semiconductor-like behavior 
($d\mu_{a}/dT<$ 0) 
above $H$ = 25 kOe. In Fig. \ref{fig:nine}(b) we show the $T$ dependence of $\mu_{a}$ 
for $H$$\parallel$$c$. The $T$ dependence of $\mu_{a}$ below 
7 K is dependent on $H$. The difference $\mu_{a}$ shows 
a metallic behavior ($d\mu_{a}/dT>$ 0) below 10 
kOe, while it shows a semiconductor-like behavior above 20 kOe. 
Note that $\mu_{a}$ at $H$ = 47.5 kOe drastically increases 
with decreasing $T$, leading to the insulating state. The magnitude 
of $\mu_{a}$ for $H$$\parallel$$c$ is on the same order as that 
for $H$$\perp$$c$. The $T$ dependence of $\rho_{a}$ 
at $H$ = 47.5 kOe for $H$$\parallel$$c$ is well described by 
Eq.(\ref{eq:three}) 
for 1.9 $\leq T\leq$ 4.3 K, where $a_{0}$ = (5.8385 
$\pm$ 0.0002) m$\Omega$cm and $a_{1}$ = (5.33 $\pm$ 
0.23) $\times$ 10$^{-3}$ m$\Omega$cm. The corresponding fitting curve 
is denoted by the solid line in the inset of Fig. \ref{fig:nine}(b). The 
ratio $a_{1}/a_{0}$ 
= 9.13 $\times$ 10$^{-4}$ is almost the same as that (= 7.85 $\times$ 10$^{-4}$) for $\rho_{a}$ 
at $H$ = 45 kOe for $H$$\perp$$c$ (see Sec. \ref{resultC}), indicating 
the occurrence of 2D WLE.

Bi-MG undergoes a transition from the metallic phase to the semiconductor-like 
phase at a critical field $H_{c}$. Similar crossover behavior 
is observed in Si MOSFET,\cite{Simonian1997} and amorphous metal films 
of InO$_{x}$,\cite{Hebard1990} 
MoGe,\cite{Mason1999} and Bi.\cite{Markovic1998} The value of $H_{c}$ ($\approx$ 25 kOe) 
for Bi-MG is almost the same as that for Si MOSFET\cite{Simonian1997} and amorphous 
MoGe film.\cite{Mason1999} Note that the Zeemann energy $gS\mu_{B}H$ 
at 25 kOe corresponds to a thermal energy $k_{B}T_{H}$ with $T_{H}$ 
= 1.7 K, where $g$ = 2 and $S$ = 1/2. The temperature $T_{H}$ 
is slightly lower than $T_{c}$ (= 2.48 K) for the superconductivity. 
The suppression of the metallic phase by $H$ is independent 
of the directions of $H$ ($H$$\parallel$$c$ and $H$$\perp$$c$) 
for Bi-MG. These results suggest that the spin related effect 
is significant compared to the orbital effect. This is consistent 
with the result derived from the susceptibility measurement that 
the Curie-like susceptibility is dominant at low $T$.

\section{CONCLUSION}
Bi-MG shows superconductivity at $T_{c}$ = 2.48 K, where the 
coherence length is much larger than the size of nanoparticles. 
The spin related effect characterized by a Curie-like susceptibility 
is enhanced, while the orbital effect is greatly suppressed. 
The transition from metallic phase to semiconductor-like phase 
is induced by the application of $H$ above 25 kOe. A negative 
magnetoresistance in $\rho_{a}$ and a logarithmic divergence 
of $\rho_{a}$ with decreasing $T$ are indicative of the 
2D WLE. 

\begin{acknowledgments}
We would like to thank Kikuo Harigaya for valuable comments on 
the magnetism of nanographites with zigzag edges. The work at 
SUNY-Binghamton was supported by SUNY-Research Foundation (Grant 
No. 240-9522A). The work at Osaka University was supported by 
the Ministry of Cultural Affairs, Education and Sport, Japan 
under the grant for young scientists (No. 70314375) and by Kansai 
Invention Center, Kyoto, Japan. We were 
grateful to Advanced Ceramics Corporation, Ohio for providing 
us with HOPG (grade ZYA). 
\end{acknowledgments}

\end{document}